# Voltage-assisted Magnetization Switching in Ultrathin $Fe_{80}Co_{20}$ Alloy Layers

Yoichi Shiota[1], Takuto Maruyama[1], Takayuki Nozaki[1,2], Teruya Shinjo[1], Masashi Shiraishi[1], Yoshishige Suzuki[1]

[1]*Graduate school of Engineering Science, Osaka University, 1-3 Machikaneyama, Toyonaka, Osaka 560-8531, Japan*

[2]*PRESTO, JST, 4-1-8 Honcho, Kawaguchi, Saitama 332-0012, Japan*

Growing demands for the voltage-driven spintronic applications with ultralow-power consumption have led to new interest in exploring the voltage-induced magnetization switching in ferromagnetic metals. In this study, we observed a large perpendicular magnetic anisotropy change in Au(001) / ultrathin $Fe_{80}Co_{20}$(001) / MgO(001) / Polyimide / ITO junctions, and succeeded in realizing a clear switching of magnetic easy axis between in-plane and perpendicular directions. Furthermore, employing a perpendicularly magnetized film, voltage-induced magnetization switching in the perpendicular direction under the assistance of magnetic fields was demonstrated. These pioneering results may open a new window of electric-field controlled spintronics devices.

Spin-polarized current induced manipulation of magnetization in nano-scale ferromagnets[1-3] is one of the promising candidates for a novel writing technique in magnetic memory devices due to the low critical current density for the magnetization switching. This technique succeeds in reducing the writing current, compared with the case of current induced magnetic field. However, it still consumes higher energy than the stabilization energy required for single-bit information. Since further reduction of the power consumption can be expected in voltage-based devices, as was proven in C-MOS technology, the use of voltage to control the magnetization would be ideal for future magnetic memories and logic elements.

In order to realize the electric-field induced manipulation of magnetization, several experimental and theoretical approaches have been made, such as magnetostriction in multilayered stacks, including piezoelectric materials,[4,5] electric-field control of ferromagnetism and magnetic anisotropy in ferromagnetic semiconductors,[6-9] multi-ferroic materials,[10] magneto-electric switching of exchange bias[11-13] and the magneto-electric interface effect.[14-17] Recently, we succeeded in achieving the voltage control of perpendicular magnetic anisotropy in Au(001) / bcc-Fe(001) / MgO(001) junctions.[18] An important point of this finding is that it can be combined with conventional MgO-based magnetic tunnel junctions,[19,20] which are already highly successful in the field of magnetic sensors for high-density HDD and magnetic random access memories.

In the previous study, we used a few atomic layers of Fe grown on a Au(001) surface. This layer possesses the perpendicular magnetic anisotropy, however the easy axis was in-plane over all thickness ranges, due to the demagnetizing field. Thus, magnetization switching could not be controlled in such a system.

Since we can control the perpendicular magnetic anisotropy by a voltage application, a voltage-induced coercivity change should be possible if we replace the Fe layer with perpendicularly magnetized materials. This leads to the voltage-assisted magnetization switching with the assisting magnetic field. A similar demonstration was reported in a ferromagnetic semiconductor system.[7] However, if we can realize it in a conventional 3d-ferromagnetic metal / MgO based junction at room temperature, it should make large impact on spintronics devices. For the work reported here, we focused on ultrathin $Fe_{80}Co_{20}$ alloy layers to obtain a perpendicularly magnetized film, and succeeded in realizing the voltage-induced magnetization switching under the assisting magnetic field.

As schematically shown in Fig.1, the sample structure stack includes an MgO substrate with MgO(10 nm) / Cr(10 nm) / Au(50 nm) / $Fe_{80}Co_{20}$($t$ : 0-0.75 nm) / MgO(10 nm) / polyimide(1,500 nm) / ITO(100 nm). We prepared multilayered epitaxial films, except for top thick polyimide and ITO layers, by using a molecular beam epitaxy method. The Au buffer layer was annealed at 200 $^o$C after deposition at room temperature to improve the surface flatness. The $Fe_{80}Co_{20}$ alloy was deposited using a co-evaporation method from Fe and Co sources. As the perpendicular surface magnetic anisotropy of $Fe_{80}Co_{20}$ layer exhibits a sensitive dependence on the thickness, we fabricated a wedge-shaped $Fe_{80}Co_{20}$ layer over a 2cm wide wafer. After depositing the top MgO layer, the sample was removed from the chamber and the surface was coated with a polyimide layer by using a spin-coater (then annealed at 200 $^o$C). An ITO (Indium tin oxide) electrode, 1mm in diameter, was fabricated using a metal mask. The bias voltage was applied between the top ITO and the bottom Au electrodes. The bias direction was defined with respect to the top ITO electrode. To investigate the magnetic hysteresis curve of ultrathin $Fe_{80}Co_{20}$ alloy, we measured the Kerr ellipticity, $\eta_k$, in a polar configuration, under the application of voltage.

Figure 2(a) shows magnetic hysteresis curves of $Fe_{80}Co_{20}$ without the voltage application. A clear transition of magnetic easy axis from in-plane to perpendicular, depending on the $Fe_{80}Co_{20}$ layer thickness, differing from our previous Au / Fe / MgO system[18], was observed at around $t = 0.55$nm. The calculations of $E_{perp}t$, together with the liner fitting[18], appears in Fig. 2(b). Here $E_{perp}$ is the perpendicular magnetic anisotropy energy per unit volume of the film and $t$ is the film thickness of $Fe_{80}Co_{20}$ layer. The results obtained in the Fe / MgO system are also shown in the same figure. From the intercept, indicating the surface anisotropy, $K_s$, of the Au / $Fe_{80}Co_{20}$ / MgO junctions, was estimated to be 650 $\mu J/m^2$. This value is greater than that of Au / Fe / MgO junctions, 580 $\mu J/m^2$.

We measured the polar-Kerr hysteresis curves with a 0.58 nm thick $Fe_{80}Co_{20}$ layer under the two bias voltage applications of ±200 V,[21] as shown in Fig.3. When we changed the voltage from +200 V to -200 V, a perpendicular magnetic anisotropy was clearly induced. It also should be noted that the magnetic easy axis was electrically manipulated between in-plane and perpendicular directions. This change with respect to the applied voltage was reversible and the sign of the anisotropy change was the same as in the previous results observed with the Au / Fe / MgO system.[18] The first-principles calculation predict that the magnetic anisotropy energy in the Fe / Au multilayer structure shows oscillatory behavior as a function of band filling.[22] It is thought that such significant change in the magnetic anisotropy with respect to the band filling and alloy composition has a relation with a relative position of the critical points in the $d$-bands, such as band edge and singular points, to the Fermi level. However first-principles calculation including the electric-field effect is required for the further understandings.[23),24)]

Figure 4 (a) shows polar-Kerr hysteresis curves of a $Fe_{80}Co_{20}$ layer with a thickness of 0.48 nm, under the voltage application of ±200 V. We can clearly see that the $Fe_{80}Co_{20}$ layer was magnetized perpendicular to the film plane in this thickness range. The coercivity could also be controlled by the voltage application. The coercivity change induced by the voltage application of ±200 V achieved about 4 Oe, indicating that the magnetization process could be assisted electrically in this system.

For example, by applying a negative saturation magnetic field of -100 Oe, followed by increasing the external magnetic field to just below the positive $H_c$ (37Oe : point A), the magnetization state settled into A state in Fig. 4 (a). This state was a stable condition under zero bias voltage (not shown here). However, it would be unstable under the positive voltage application due to the decrease in the perpendicular magnetic anisotropy, leading to the magnetization switching from A state to B state. Once the magnetization was switched, the magnetization remained that state, even if we applied a negative voltage which should increase the coercivity. Then, after decreasing the external magnetic field from point B to just below the negative $H_c$ (-35Oe ; point C), after setting the bias voltage to zero, the reverse process of the switching from C state to D state could be induced by a positive voltage application. Figure 4 (b) shows the demonstration of voltage-induced switching using the above sequence. Sharp switching of magnetization from A state to B state and from C state to D state was induced by the voltage application at around +10 V and +5 V, respectively. Note that this electrical magnetization switching, assisted by the small bias voltage, is a reversible phenomenon in the presence of the appropriate external magnetic fields.

In summary, we have demonstrated voltage-assisted magnetization switching in the Au(001) / bcc-Fe$_{80}$Co$_{20}$(001) / MgO(001) / Polyimide / ITO junctions. The present results suggest that the perpendicular magnetization of Fe$_{80}$Co$_{20}$ layer, sandwiched between Au(001) and MgO(001), can be electrically manipulated under a small assisting magnetic field. This may thus provide a significant development of novel electric-field controlled magnetic devices.

**Acknowledgment**   We would like to thank T. Shoji, D. Yamaguchi, T. Sada, Y. Sobajima, T. Toyama and H. Okamoto for their assistance in ITO deposition. We also acknowledge S. –S. Ha, C. –Y. You for their fruitful discussions. Part of the research was conducted under the financial support of Grant-in-Aid for Scientific Research (A19206002) and G-COE program of Ministry of Education, Culture, Sports, Science and Technology-Japan (MEXT).

**Fig. 1** The sample structure used for voltage-induced magnetization switching. A positive bias voltage is defined as the positive voltage on top ITO electrode with respect to the bottom Au electrode.

**Fig. 2** (a) The thickness dependence of magnetic hysteresis curves of the ultrathin $Fe_{80}Co_{20}$ layers sandwiched between the Au(001) and MgO(001) layers. (b) Fe[18] (open circles) and $Fe_{80}Co_{20}$ (solid circles) layer thickness dependence of $E_{perp}t$ at zero bias voltage.

**Fig. 3** Magnetic hysteresis curves of a 0.58nm thick $Fe_{80}Co_{20}$ layer, measured under positive (blue) and negative (red) bias voltage applications.

**Fig. 4** (a) Hysteresis curves of a 0.48nm thick $Fe_{80}Co_{20}$ layer under the voltage application of 200V (blue) and -200V (red). (b) Kerr ellipticity, $\eta_k$, intensity as a function of applied bias voltage. Clear transition between two magnetic states was observed under the presence of adequate assisting magnetic fields.

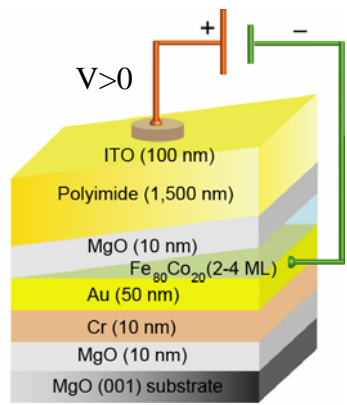

**Fig. 1**

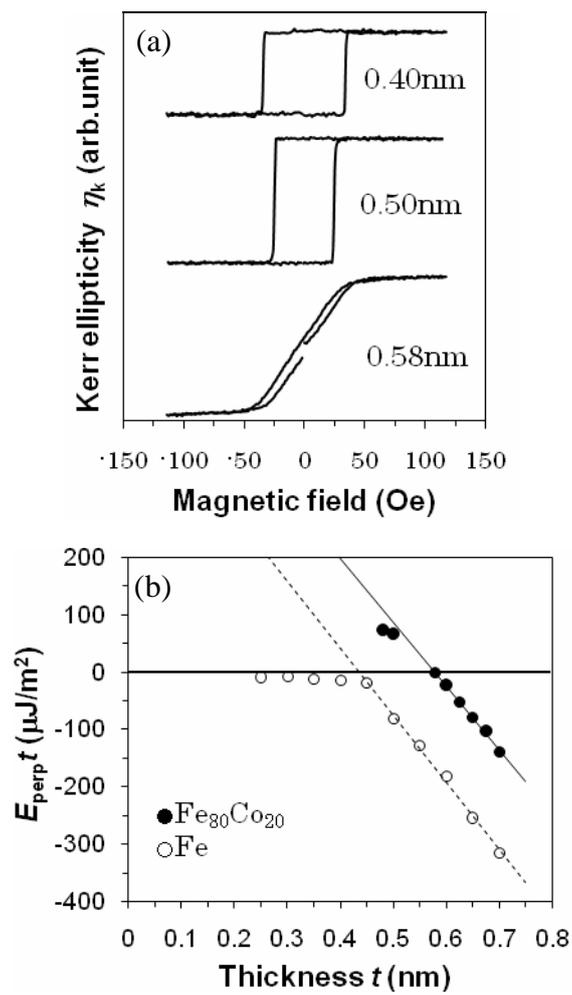

**Fig. 2**

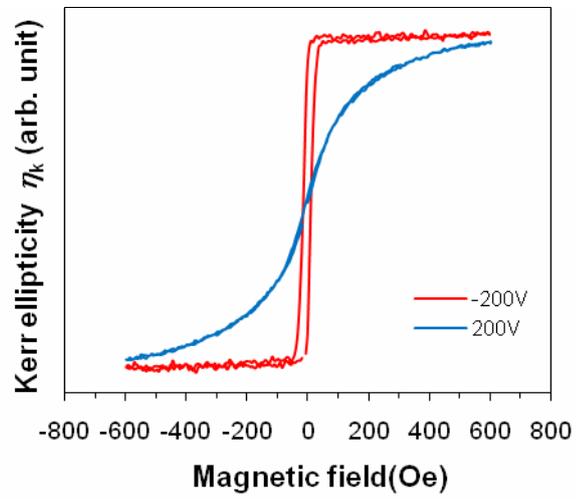

**Fig. 3**

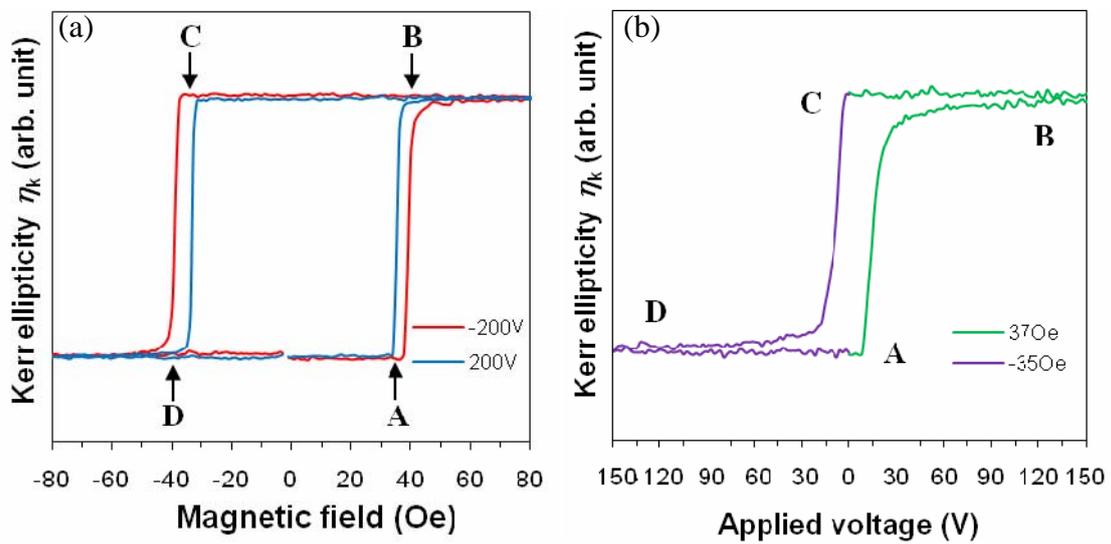

**Fig. 4**